\documentstyle[a4,12pt,amssymb,float,epsfig,psfig,times]{article}
\newcommand{\be}{\begin{equation}}
\newcommand{\ee}{\end{equation}}
\newcommand{\bea}{\begin{eqnarray}}
\newcommand{\eea}{\end{eqnarray}}

\def\p1{\pi_1}

\def\l{\lambda}
\def\f{\phi}
\def\k{\kappa}
\def\r{\rho}

\def\pmas{\partial_+}
\def\pmen{\partial_-}
\def\v{\vskip}
\renewcommand{\thefootnote}{\fnsymbol{footnote}}

\begin{document}

\begin{titlepage}
\vspace*{\stretch{0}}
\begin{flushright}
{\tt FTUV-00-0605\\
     IFIC/00-34\\
     hep-th/0006035}
\end{flushright}

\vspace*{0.5cm}

\begin{center}
{\Large\bf Quantum evolution of near-extremal Reissner-Nordstr\"om
black holes}
\\[0.5cm]
A. Fabbri$^{\rm a}$\footnote{\tt fabbria@bo.infn.it},
D. J. Navarro$^{\rm b}$\footnote{\tt dnavarro@ific.uv.es} and
J. Navarro-Salas$^{\rm b}$\footnote{\tt jnavarro@ific.uv.es}
\\[0.5cm]
{\footnotesize
a) Dipartimento di Fisica dell'Universit\`a di Bologna and INFN
sezione di Bologna,\\
Via Irnerio 46, 40126 Bologna, Italy.
\\[0.5cm]
b) Departamento de F\'{\i}sica Te\'orica and IFIC, Centro Mixto Universidad
de Valencia-CSIC.\\
Facultad de F\'{\i}sica, Universidad de Valencia, Burjassot-46100, Valencia,
Spain.}
\end{center}
\bigskip
\begin{abstract}
We study the near-horizon AdS$_2\times$S$^2$ geometry of evaporating
near-extre\-mal Reissner-Nordstr\"om black holes interacting with null matter.
The non-local (boundary) terms $t_{\pm}$, coming from the effective theory
corrected with the quantum Polyakov-Liouville action, are treated as dynamical
variables. We describe analytically the evaporation process which turns out to
be compatible with the third law of thermodynamics, i.e., an infinite amount
of time is required for the black hole to decay to extremality. Finally we
comment briefly on the implications of our results for the information loss
problem.
\end{abstract}
\vspace*{\stretch{1}}
\begin{flushleft}
PACS number(s): 04.70.Dy, 04.62.+v
\end{flushleft}
\vspace*{\stretch{0}}
\end{titlepage}
\newpage

\renewcommand{\thefootnote}{\arabic{footnote}}
\setcounter{footnote}{0}

\section{Introduction}

Since Hawking discovered that black holes decay by emission of thermal
radiation \cite{h1}, a lot of work has been done on quantum aspects of black
holes aiming to improve our understanding of the quantization of the
gravitational field. Hawking's result provided a physical meaning
to the formal analogy between the classical laws of black hole dynamics and
those of thermodynamics. But, in turn, it has been the origin of some
intriguing puzzles. Indeed, it was Hawking himself the first who pointed out
that the information can be lost as a pure quantum state collapses
gravitationally into a black hole which then evaporates into a mixed state
\cite{h2}. An opposite scheme has been suggested by 't Hooft \cite{th} to
maintain quantum coherence. Since then, several alternatives have been
proposed (see the reviews \cite{ppb}) in order to avoid the information loss
although the problem is still
unsolved. A final answer requires a complete and consistent theory of quantum
gravity. Despite the recent achievements of String Theory to explain the
Bekenstein-Hawking formula for extremal and near-extremal black holes
\cite{sv,cm,ms}, this theory is still far to describe the evaporation process
of a
black hole. However, restricting the situation to the scattering of low-energy
particles with zero angular momentum off extremal black holes one can get a
more tractable problem. This was the main reason to analyse dilatonic black
holes \cite{ghs,cghs}. The problem can be reduced then to study a
two-dimensional effective theory which turns out to be solvable
\cite{rst1,rst2,bpp,mik1,mik2,fr,cn}. However, the dilatonic black holes have
very special
properties which make it hard to extrapolate the physical picture of their
evaporation process. Near extremality they have a constant temperature, so the
decay of near-extremal black holes is governed by a Hawking radiation with a
mass independent temperature. Although this makes the mathematical treatment
of the non-locality of the effective action easier, the general situation
is more involved. In fact, for near-extremal Reissner-Nordstr\"om (RN) black
holes, the temperature depends on the mass and this makes elusive an exact
analytical framework. The aim of this paper is to show that one can get exact
quantum results of the dynamical evolution of near-extremal RN black holes in a
region very close to the horizon.\\

The paper is organised as follows. In section 2 we shall show that
near-extremal RN black holes, in the S-wave approximation and near the horizon,
can be described by the Jackiw-Teitelboim (JT) model \cite{jt}. In section
3 we shall consider the formation of a near-extremal black hole by throwing a
low-energy neutral particle into an extremal one and, in section 4, we shall
describe the Hawking radiation within the near-horizon scheme. In section 5 we
shall analyse the back reaction effects obtained when the classical equations
are modified with terms proportional to $\hbar$ coming from the
effective Liouville-Polyakov action. We find that a near-extremal black hole
evaporates within a finite proper time and after the end-point, the evaporating
solution exactly and smoothly matches with the extremal black hole geometry.
Therefore one could expect that extremal RN black holes are really stable
end-points of Hawking evaporation. However a further quantum correction changes
this picture and suggests that only an exact treatment of back reaction can
produce a reliable result. In section 6 we shall provide the exact results,
outlined in \cite{fnn}, coming from equations admitting a series
expansion in powers of $\hbar$. We find that the evaporation process requires
an infinite amount of time, in agreement with previous results based on the
adiabatic approximation \cite{st}. Finally, in section 7, we shall state our
conclusions and comment on the implications of our results for the information
loss problem.

\setcounter{equation}{0}
\section{Reissner-Nordstr\"om black holes near extremality and 
Jackiw-Teitelboim theory}

The RN black hole has been widely studied in literature. Recently, its interest
has increased as an example of AdS$_2$ arising as a near-horizon geometry. In
this section we shall review some of these features and show the close
connection with Jackiw-Teitelboim (JT) theory. The RN metric is given by
\be
\label{rn1}
d\bar{s}^2 = -U(r) dt^2 + \frac{dr^2}{U(r)} + r^2 d\Omega^2 \, ,
\ee
where
\be
\label{ur}
U(r) = \frac{(r-r_+)(r-r_-)}{r^2} = 1 - \frac{2l^2m}{r} +
\frac{l^2q^2}{r^2} \, ,
\ee
$l^2$ is Newton's constant and
\be
\label{horizons}
r_{\pm} = l^2 m \pm l\sqrt{l^2m^2 - q^2} \, ,
\ee
are the two roots of $U(r)$. Taking into account (\ref{horizons}), there are
three cases to consider: for $lm<|q|$ there is no horizon and the singularity
at $r=0$ is naked\footnote{This case is similar to the one $m<0$ of the
Schwarzschild black hole.}. For $lm>|q|$, $r_{\pm}$ are the respective inner
($r_-$) and outer ($r_+$) horizons and the thermodynamical variables, entropy
and temperature, associated to the outer horizon are given by
\be
S = \frac{\pi r_+^2}{l^2} \, , \qquad T_H = \frac{\k_+}{2\pi} \, ,
\ee
where
\be
\label{surfgravity}
\k_+ = \frac{r_+-r_-}{2r_+^2} \, ,
\ee
is the surface gravity on the outer (event) horizon.
Finally, for the critical value of the mass $m=m_0=l^{-1}|q|$, both inner and
outer horizons merge to $r_0=l^2m_0=l|q|$. This is the extremal black
hole with a vanishing temperature.\\

Now we consider small perturbations near extremality $m=m_0+\Delta m$, keeping
the charge $q$ unchanged. The inner and outer horizons (\ref{horizons}), to
leading order in $\sqrt{\Delta m}$ read
\be
\label{nehorizons}
r_{\pm} = lq \pm l \sqrt{2ql\Delta m} \, ,
\ee
where from now on we assume $q=|q|$. The extremal black hole is recovered for
$\Delta m=0$. Near extremality the entropy deviation $\Delta S=S_H-S_0$ and
temperature are given by
\bea
\label{neentropy}
\Delta S &=& 4\pi \sqrt{\frac{lq^3\Delta m}{2}} \, , \\
\label{netemperature}
T_H &=& \frac{1}{2\pi} \sqrt{\frac{2\Delta m}{lq^3}} \, .
\eea

The RN metric (\ref{rn1}) comes from the Einstein-Maxwell action
\be
I = \frac{1}{16\pi l^2} \int d^4x \sqrt{-\bar{g}^{(4)}} 
(\bar{R}^{(4)} - l^2 (\bar{F}^{(4)})^2) \, .
\ee
Assuming spherical symmetry, dimensional reduction leads to the
following effective two-dimensional theory
\be
\label{effective}
I = \int d^2x \sqrt{-g} (R\f + l^{-2} V(\f)) \, .
\ee
To get the above expression we have performed a conformal reparametrization
of the metric 
\be
\label{rescaled}
ds^2=\sqrt{\f} d\bar{s}^2 \, ,
\ee
where
\be
\label{coordinate}
\f=\frac{r^2}{4l^2} \, ,
\ee
and
\be
V(\f) = (4\f)^{-\frac{1}{2}} - q^2(4\phi)^{-\frac{3}{2}} \, .
\ee
The extremal configuration is recovered for the zero of the potential
$V(\f_0)=0$. Moreover expanding (\ref{effective}) around $\f_0=\frac{q^2}{4}$
\bea
\label{e1}
\f = \f_0 + \tilde{\f} \, , \\
\label{e2}
m = m_0 + \Delta m \, ,
\eea
we get
\be
\label{jtaction}
I = \int d^2x \sqrt{-g} (R \tilde{\f} + l^{-2} \tilde{V}(\tilde{\f})) +
{\cal{O}}(\tilde{\f}^2) \, ,
\ee
where $\tilde{V}(\tilde{\f})$ is given by
\be
\tilde{V}(\tilde{\f}) =  V^{\prime}(\f_0) \tilde{\f} = \frac{4}{q^3} 
\tilde{\f} \, ,
\ee
and the leading order term is just the JT theory. We can also
get this result studying the behavior of the metric (\ref{rescaled}) near
extremality. The general solution of (\ref{effective}) is \cite{gkm}
\bea
\label{m1}
ds^2 &=& -(J(\f)-lm) dt^2 + (J(\f)-lm)^{-1} dx^2 \, , \\
\label{m2}
\tilde{\f} &=& \frac{x}{l} \, ,
\eea
where $J(\f)=\int^{\f} d\bar{\f} V(\bar{\f})$. The thermodynamical variables,
in terms of the two-dimensional effective theory, read \cite{mk}
\bea
S &=& 4\pi \f_+ \, , \\
T_H &=& \frac{1}{4\pi} \left| \frac{dU(r)}{dr} \right|_{r_+} = \frac{1}{2\pi}
\frac{V(\f_+)}{2l} \, ,
\eea
where $J(\f_{\pm})-l\Delta m=0$ and we take into account that
\be
\sqrt{\f}U(r(\f)) = J(\f) - lm \, .
\ee

Expanding (\ref{m1}), (\ref{m2}) around $\f_0$ we get\footnote{A slightly
different near-horizon approach, in which the expansion is taken around the
outer horizon $r_+$, has been considered in \cite{mms,ss}}
\bea
\label{jtmetric}
ds^2 &=& -(\tilde{J}(\tilde{\f})-l\Delta m) dt^2 + (\tilde{J}(\tilde{\f})-
l\Delta m)^{-1} d\tilde{x}^2 + {\cal{O}}(\tilde{\f}^3) \, , \\
\tilde{\f} &=& \frac{\tilde{x}}{l} \, ,
\eea
where
\be
\tilde{J}(\tilde{\f}) = \frac{2}{q^3} \tilde{\f}^2 \, .
\ee
The leading order terms in the above expansion are AdS$_2$ geometries, which
are the solutions of the JT theory. Moreover, the mass deviation $\Delta m$ is
just the conserved parameter of JT theory
\be
\label{adm}
l \Delta m = \tilde{J}(\tilde{\f}) - l^2 |\nabla \tilde{\f} |^2 \, .
\ee
Therefore the JT theory describes both extremal ($\Delta m=0$) and
near-extremal RN black holes.\\

Let us now see how the deviations from extremality of the thermodynamical
variables are given in terms of JT magnitudes. After the near-horizon
approximation (\ref{e1}), (\ref{e2}), we get
\bea
\label{jtentropy}
\Delta S &=& 4\pi \tilde{\f}_+ \, , \\
\label{jttemperature}
T_H &=& \frac{1}{2\pi} \frac{\tilde{V}(\tilde{\f})}{2l} \, ,
\eea
where $\tilde{\f}_+$ is the positive root of $J(\tilde{\f})-l\Delta m=0$
\be
\tilde{\f}_+ = \sqrt{\frac{ql\Delta m}{2}} \, ,
\ee
and for the above value, (\ref{jtentropy}), (\ref{jttemperature}) coincide with
the near-extremal values (\ref{neentropy}), (\ref{netemperature}). This is
nothing but JT thermodynamics\footnote{A realisation of the AdS$_2$/CFT$_1$
correspondence in the JT theory accounts for the deviation from extremality of
the Bekenstein-Hawking entropy of RN black holes \cite{nnn1,nnn2}.} All these
features can be resumed in the following table\\
 
\begin{center}
\begin{tabular}{|l|l|}
\hline
Near-extremal black hole & Jackiw-Teitelboim theory \\ \hline
& \\
radius deviation $r-r_0$ & $\frac{2l}{q}\tilde{\f}$ \\
mass deviation $\Delta m$ & JT mass $\tilde{m}$ \\
inner and outer horizons & AdS$_2$ horizons \\
entropy deviation $\Delta S$ & $S_{JT}$ \\
$T_H$ & $T_{JT}$ \\ \hline
\end{tabular}
\end{center}

\v1cm

To finish this section we review some aspects of the matter-coupled theory
given by the action
\be
\label{jtactionm}
\int d^2x \sqrt{-g} \left( R \tilde{\f} + 4\l^2 \tilde{\f} -
\frac{1}{2} |\nabla f|^2 \right) \, ,
\ee
where the field $f$ models the matter degrees of freedom. Note that the field
$f$ propagates freely as it happens in the original 4d theory in a region very
close to the horizon. In conformal gauge $ds^2=-e^{2\r}dx^+dx^-$, the equations
of motion derived from the above action are
\bea
\label{eq1}
2\pmas \pmen \r + \l^2 e^{2\r} &=& 0 \, , \\
\label{eq2}
\pmas \pmen \tilde{\f} + \l^2 \tilde{\f} e^{2\r} &=& 0 \, , \\
\label{eq3}
\pmas \pmen f &=& 0 \, , \\
\label{eq4}
-2\partial^2_{\pm} \tilde{\f} + 4 \partial_{\pm} \rho \partial_{\pm} 
\tilde{\f} - T^f_{\pm \pm} &=& 0 \, ,
\eea
where $T^f_{\pm \pm}= (\partial_{\pm} f)^2$ is the stress tensor of matter
fields. The general solution can be written in terms of four chiral
functions $A_{\pm}(x^{\pm})$, $a_{\pm}(x^{\pm})$ \cite{cinn} \cite{f}
\bea
\label{jtconf1}
ds^2 &=& -\frac{\pmas A_+ \pmen A_-}{(1+\frac{\lambda^2}{2} A_+ A_-)^2}
dx^+ dx^- \, , \\
\label{jtconf2}
\tilde{\phi} &=& -\frac{1}{2} \left( \frac{\pmas a_+}{\pmas A_+} +
\frac{\pmen a_-}{\pmen A_-} \right) + \frac{\lambda^2}{2} 
\frac{A_+ a_- + A_- a_+}{1+\frac{\lambda^2}{2} A_+ A_-} \, ,
\eea
obeying the constrain equations
\be
\label{gc}
\partial_{\pm}^2 \left( \frac{\partial_{\pm} a_{\pm}}
{\partial_{\pm} A_{\pm}} \right) -
\frac{\partial_{\pm}^2 A_{\pm}}{\partial_{\pm} A_{\pm}} \;
\partial_{\pm} \left( \frac{\partial_{\pm} a_{\pm}}{\partial_{\pm}
A_{\pm}} \right) = T^f_{\pm \pm} \, .
\ee
In the absence of matter fields, these solutions are parametrized by a
diffeomorphism invariant quantity, related with the mass, which for this case
reads (see (\ref{adm}))
\be
\label{localmass}
\tilde{m} = \frac{2}{lq^3} \tilde{\f}^2 - l^2 | \nabla \tilde{\f} |^2 \, .
\ee
When $T^f_{--}=0$, $\tilde{m}$ is a chiral function ($\pmen \tilde{m}$=0)
having the physical meaning of local mass \cite{cfn2}
\be
\label{econservation}
\tilde{m}(x^+) = \tilde{m}_0 -2l^2 \int dx^+ e^{-2\r} \pmen \tilde{\f} \,
T^f_{++} \, ,
\ee
where $\tilde{m}_0$ is the primordial mass of the black hole in the absence of
matter fields.\\

Finally, we would like to stress the fact that the four chiral functions
$A_{\pm}(x^{\pm})$, $a_{\pm}(x^{\pm})$ define two free fields with a quadratic
stress-tensor equals to the left hand side of (\ref{gc}) \cite{cinn}. Moreover
the transformation from the gravity variables to the free fields is a canonical
transformation which makes the underlying conformal symmetry of the theory
more transparent.

\setcounter{equation}{0}
\section{Making a near-extremal black hole}

In this section we shall study the process which makes a black hole leave
extremality due to infalling null neutral matter. This can be done
analytically by means of a Vaidya-type metric
\be
\label{rn2}
d\bar{s}^2 = -\left( 1 - \frac{2l^2m(v)}{r} + \frac{l^2q^2}{r^2} \right) dv^2
+ 2 dr dv + r^2 d \Omega^2 \, ,
\ee
generated by the following stress tensor for the infalling matter
\be
\bar{T}_{vv}^{(4)}=\frac{\partial_v m(v)}{4\pi r^2} \, .
\ee
It is possible to match an extremal black hole solution with a near-extremal
one by means of a shock wave along the line $v=v_0$. The corresponding mass
function is given by
\be
m(v) = l^{-1}q + \Delta m \Theta(v-v_0) \, ,
\ee
where $\Theta$ is the step function and the stress tensor is
\be
\bar{T}_{vv}^{(4)} = \frac{\Delta m}{4\pi r^2} \delta(v-v_0) \, .
\ee
In terms of the two-dimensional effective theory (\ref{jtaction}), where
$\l^2=l^{-2}q^{-3}$, the stress tensor of matter fields read
\be
T^f_{\mu\nu} = 4\pi r^2 \bar{T}_{\mu\nu}^{(4)} \, ,
\ee
and the shock wave turns into
\be
\label{shock}
T^f_{vv} = \Delta m \delta(v-v_0) \, .
\ee

Now we go back to the metric (\ref{rn2}) and make use of the near-horizon
approximation considered in previous section. After the conformal
reparametrization (\ref{rescaled}) and performing the expansion (\ref{e1}),
(\ref{e2}), the metric (\ref{rn2}) becomes
\be
\label{rn2h}
ds^2 = -\left( \frac{2\tilde{x}^2}{l^2q^3}-l\Delta m \Theta(v-v_0)\right) dv^2
+ 2d\tilde{x} dv \, ,
\ee
where $\tilde{x}=l\tilde{\f}$. The Penrose diagram of this process is
represented in Fig.1
\begin{figure}[H]
\centerline{\psfig{figure=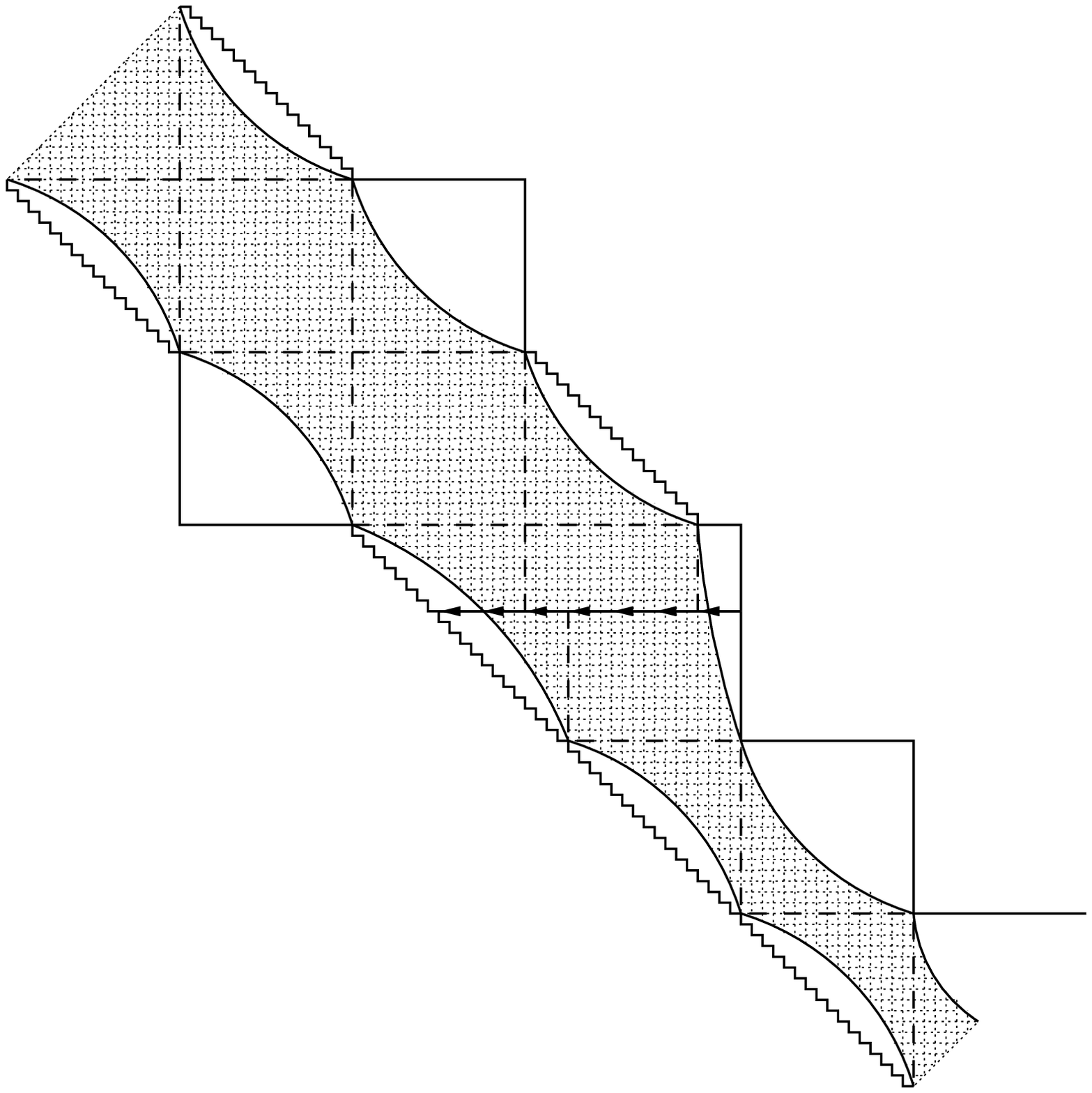,width=3.in,angle=-45}}
\end{figure}
\begin{center}
\makebox[12cm]{\parbox{12cm} {\small Fig.1. Penrose diagram corresponding to
the creation of a near-extremal RN black hole from the extremal one. Dashed
lines are the black hole horizons and the shaded strip is the near-horizon
AdS$_2$ geometry. The arrow line represents the infalling shock wave. The
shaded region between the AdS$_2$ boundaries is described by Kruskal
coordinates in Fig.3.}}
\end{center}
\v1cm

In terms of the coordinates $x^{\pm}$ defined by
\bea
\label{tr1}
x^+ &=& v \, , \\
\label{tr2}
x^- &=& v + \frac{l^2q^3}{\tilde{x}} \, ,
\eea
for $v<v_0$, and
\bea
\label{tr3}
v &=& x_0^+ + \sqrt{\frac{2lq^3}{\Delta m}} {\mathrm arctanh}
\sqrt{\frac{\Delta m}{2lq^3}} (x^+-x_0^+) \, , \\
\label{phi}
\tilde{x} &=& lq^3 \frac{1-\frac{\Delta m}{2lq^3} (x^+-x_0^+)(x^--x_0^+)}
{x^--x^+} \, ,
\eea
for $v>v_0$ (where $x_0^+=v_0$) the metric (\ref{rn2h}) turns out
\be
\label{jt1}
ds^2 = - \frac{2l^2q^3 dx^+ dx^-}{(x^--x^+)^2} \, ,
\ee
which corresponds to the solution (\ref{jtconf1}) of JT theory with the
following gauge fixing
\bea
\label{gauge+}
A_+ &=& x^+ \, , \\
\label{gauge-}
A_- &=& \frac{-2}{\l^2 x^-} \, ,
\eea
where $\l^2=l^{-2}q^{-3}$. The corresponding dilaton functions are
\bea
\label{jt2}
\tilde{\f} &=& \frac{lq^3}{x^--x^+} \, , \qquad v<v_0 \, ,\\
\label{jt3}
\tilde{\f} &=& lq^3 \frac{1-\frac{\Delta m}{2lq^3} (x^+-x_0^+)
(x^--x_0^+)}{x^--x^+} \, , \qquad v>v_0 \, ,
\eea
which are respectively recovered from (\ref{jtconf2}) when
\be
a_+ = -lq^3 \, , \qquad a_- = 0\, ,
\ee
for $v<v_0$ and
\be
\label{aclass}
a_+ = -\frac{1}{2} x_0^+ \Delta m (x^+ + \Delta^+) \, , \qquad
a_- = -\frac{1}{2} x_0^+ \Delta m (\frac{-2}{\l^2x^-} + \Delta^-) \, ,
\ee
for $v>v_0$, where
\bea
\label{d+}
\Delta^+ &=& -x_0^+ + \frac{2}{l\Delta m \l^2 x_0^+} \, , \\
\label{d-}
\Delta^- &=& \frac{2}{\l^2 x_0^+} \, .
\eea
We see explicitly how the JT theory describes both
extremal and near-extremal geometries, being the horizon structure described
by the dilatonic function $\tilde{\f}$. It is worth to remark that the
near-extremal configuration (\ref{jt3}) leads to the extremal one (\ref{jt2})
in the limit $\Delta m=0$ and that both dilatonic functions match continuously
along $x^+=x_0^+$. So matching the discontinuity of $-2\partial^2_+ \tilde{\f}
+ 4\partial_+ \rho \partial_+ \tilde{\f}$ along $x_0^+$, we recover the shock
wave (\ref{shock}).\\

Finally, let us consider the description of the near-extremal black hole in
the near-horizon approximation. The outer and inner
horizons $r_{\pm}$ are given in $x^{\pm}$ coordinates by the curves
\be
\tilde{\f} = \tilde{\f}_{\pm} = \pm \sqrt{\frac{q^3l\Delta m}{2}} \, ,
\ee
which is equivalent to the standard condition $\partial_+ \tilde{\f}=0$ of
two-dimensional dilaton gravity. We get
\be
x^- = x_0^+ \pm a \, ,
\ee
where
\be
a = \sqrt{\frac{2lq^3}{\Delta m}} \, .
\ee

\v1cm
\begin{figure}[H]
\centerline{\psfig{figure=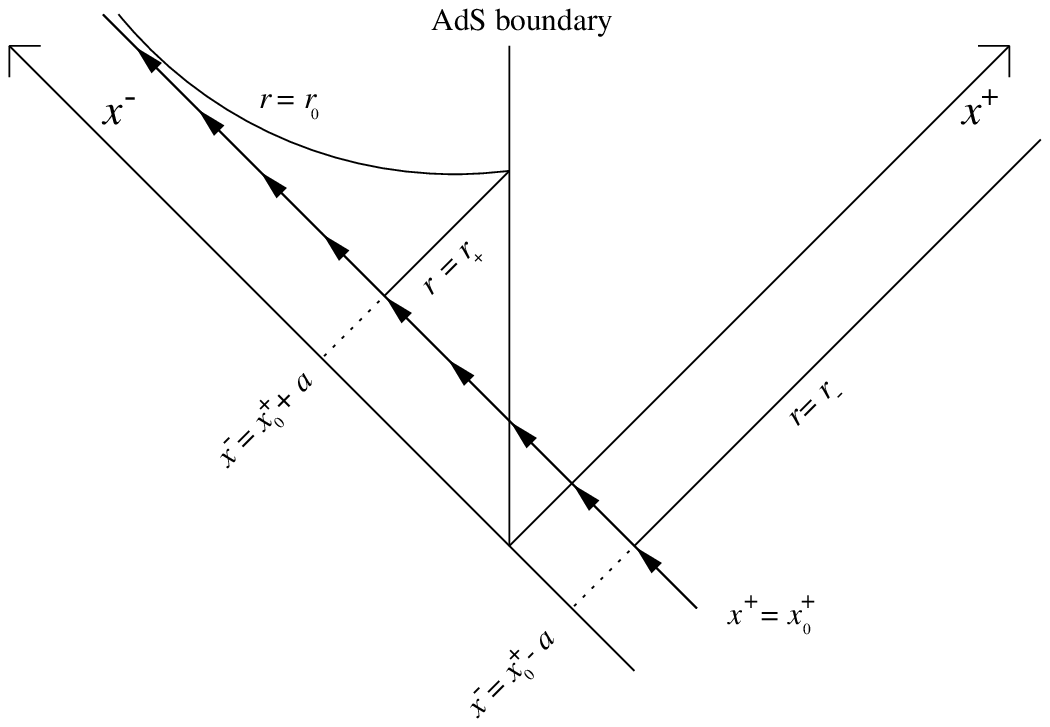,width=3.in,angle=-90}}
\end{figure}
\begin{center}
\makebox[12cm]{\parbox{12cm}{\small \noindent Fig.2. Kruskal diagram of
near-extremal RN black hole. The two timelike boundaries of near-horizon
geometry AdS$_2$ are represented by the vertical line $x^-=x^+$. The infalling
shock wave emerges from one boundary (left side of $x^-=x^+$ line) and,
crossing the outer and inner horizons, reaches the other boundary (right side
of $x^-=x^+$ line).}}
\end{center}
\v1cm

Moreover we are also interested on the curve $r=r_0$ (the old horizon
of the extremal black hole which is no longer a horizon) when the
near-extremal black hole is created. In
$x^{\pm}$ coordinates, this curve is given by $\tilde{\f}=0$ (see (\ref{e1}))
\be
(x^+-x_0^+)(x^--x_0^+) = a^2 \, .
\ee
The corresponding Kruskal diagram is given in Fig.2. As illustrated in this
figure, the two timelike boundaries of AdS$_2$ are representated in
coordinates $x^{\pm}$ by the same curve $x^-=x^+$.

\setcounter{equation}{0}
\section{Hawking radiation}

We shall focus now on the presence of Hawking's radiation in this model. To
obtain the Hawking radiation we start writing the metric (\ref{rn2}) in
light-cone coordinates $(v,u)$
\be
\label{rn3}
d\bar{s}^2 = -U(r) du dv + r^2 d\Omega^2 \, ,
\ee
where $U(r)$ is given by (\ref{ur}) and
\bea
\label{t}
\frac{v+u}{2} &=& t \, , \\
\label{r}
\frac{v-u}{2} &=& r^{\ast} \, ,
\eea
where $r^{\ast}$ is the tortoise coordinate
\be
r^{\ast} = \int \frac{dr}{U(r)} = r + \frac{1}{2\k_+} \ln |r-r_+| -
\frac{1}{2\k_-} \ln |r-r_-| \, .
\ee
After the conformal reparametrization (\ref{rescaled}) we get
\be
ds^2 = -(J(\f)-lm) du dv \, ,
\ee
where now
\be
u = v - 2l \int \frac{d\f}{J(\f)-lm} \, .
\ee
Performing the `near-horizon' approximation (\ref{e1}), (\ref{e2}), the above
expression turns out
\be
\label{uv}
u = v - 2 \int \frac{d\tilde{x}}{2\l^2\tilde{x}^2-l\Delta m
\Theta(v-v_0)} \, .
\ee
The extremal black hole before the shock wave ($v<v_0$) is given by
\be
\label{v-}
ds^2 = -2\l^2\tilde{x}^2 du dv \, ,
\ee
and the near-extremal black hole created after the shock wave ($v>v_0$)
\be
\label{v+}
ds^2 = -(2\l^2\tilde{x}^2-l\Delta m) d\bar{u} dv \, .
\ee
Integrating (\ref{uv}) for both cases we get the following relations for the
coordinate $u$ before and after the shock wave
\bea
\label{u-}
u &=& v + \frac{1}{\l^2 \tilde{x}} \, , \\
\label{u+}
\bar{u} &=& v + \sqrt{\frac{2}{l\Delta m \l^2}} {\mathrm arctanh}
\sqrt{\frac{2\l^2}{l\Delta m}} \tilde{x} \, .
\eea
Imposing the continuity of solutions (\ref{v-}) and (\ref{v+}) along $v=v_0$
(and taking also into account $\tilde{x}(v_0,u)=\tilde{x}(v_0,\bar{u})$) we
obtain the relation between the coordinate $u$ before and after the shock wave
\be
\label{ubaru}
u = v_0 + a \> {\mathrm cotanh} \frac{\bar{u}-v_0}{a} \, .
\ee
From this relation we can work out immediately the outgoing energy flux
measured by an observer near the horizon in terms of the Schwarzian derivative
between the coordinates $u$ and $\bar{u}$
\be
\langle T^f_{\bar{u}\bar{u}} \rangle = -\frac{\hbar}{24\pi} \{ u, \bar{u} \} =
\frac{\hbar}{12\pi a^2} \, .
\ee
We observe that this flux is constant and coincides with thermal value of
Hawking flux for near-extremal RN black holes
\be
\langle T^f_{\bar{u}\bar{u}} \rangle = \frac{\pi}{12} T_H^2 =
\frac{\hbar \Delta m}{24\pi lq^3} \, ,
\ee
where $T_H$ is Hawking's temperature (\ref{netemperature}). This fact can be
understood easily since
the AdS$_2\times$S$^2$ geometries associated to (\ref{v-}) and (\ref{v+})
represent indeed the near-horizon geometry of the RN geometries (\ref{rn2})
due to the shock wave (\ref{shock}). So the constant thermal flux for every
value of $\bar{u}$ corresponds to the flux measured by an inertial observer
at future null infinity approaching the event horizon of the RN black hole.

\setcounter{equation}{0}
\section{Back reaction to leading order in $\hbar$}

In this section we shall start our analysis of the back reaction of
near-extremal black holes in the near-horizon effective theory. The starting
point must be the matter-coupled classical action (\ref{jtaction}) corrected
with the Polyakov-Liouville term \cite{p}
\bea
\label{poly}
\tilde{I}_{{\mathrm eff}} &=& \int d^2x \sqrt{-g} \left(R \tilde{\phi} + 4 
\lambda^2 \tilde{\phi} -\frac{1}{2} \sum_{i=1}^N |\nabla f_i|^2\right) - 
\frac{N\hbar}{96\pi} \int \sqrt{-g} R \; \square^{-1} R \nonumber \\
&+& \xi \frac{N\hbar}{12\pi} \int d^2x \sqrt{-g} \lambda^2 \, ,
\eea
where we have considered the presence of $N$ scalar fields $f_i$ in order to
define the semiclassical theory in the large $N$ limit. Nevertheless we have
to point out the important fact that, for the JT theory, the exact quantum
effective action is locally equivalent to the one-loop corrected theory
\cite{kum} and, therefore, we could maintain $N=1$.\\

The Polyakov-Liouville action in (\ref{poly}) has a cosmological term \cite{p}
and we shall fix it in such a way that the extremal black hole solution remains
a solution of the quantum theory. This is, in some sense, analogous to the
manner in which the local counterterm of the RST model \cite{rst1,rst2} is
introduced. This argument implies that $\xi=1$. Therefore the unconstrained
equations remain as the classical ones
\bea
\label{qeq1}
2\pmas \pmen \r + \l^2 e^{2\r} &=& 0 \, , \\
\label{qeq2}
\pmas \pmen \tilde{\f} + \l^2 \tilde{\f} e^{2\r} &=& 0 \, , \\
\label{qeq3}
\pmas \pmen f_i &=& 0 \, ,
\eea
so the solutions (\ref{jtconf1}), (\ref{jtconf2}) are not modified. In
contrast, the constrain equations (\ref{eq4}) get modified according to
\be
\label{qeq4}
-2\partial^2_{\pm} \tilde{\f} + 4 \partial_{\pm} \rho \partial_{\pm} 
\tilde{\f} = T^f_{\pm \pm} - \frac{N\hbar}{12\pi} \left( (\partial_{\pm} \r )^2
-\partial_{\pm}^2 \r + t_{\pm} \right) \, ,
\ee
where the functions $t_{\pm}(x^{\pm})$ are related with the boundary
conditions of the theory and depend on the quantum state of the system. They
transform according the Schwarzian derivative to make covariant the equation
(\ref{qeq4}). In terms of the four chiral functions appearing in
(\ref{jtconf1}), (\ref{jtconf2}), the above constraints read
\be
\label{const}
\partial_{\pm}^2 \left( \frac{\partial_{\pm} a_{\pm}}
{\partial_{\pm} A_{\pm}} \right) -
\frac{\partial_{\pm}^2 A_{\pm}}{\partial_{\pm} A_{\pm}} \;
\partial_{\pm} \left( \frac{\partial_{\pm} a_{\pm}}{\partial_{\pm}
A_{\pm}} \right) = T^f_{\pm\pm} + \frac{N\hbar}{12\pi} \left(
\frac{1}{2} \{ A_{\pm}, x^{\pm} \} - t_{\pm} \right) \, .
\ee
The crucial point to go on the analysis is to choose the adequate functions
$t_{\pm}(x^{\pm})$. For the CGHS theory \cite{rst1,rst2,bpp} the correct
choice is $t_{\pm}(x^{\pm})=\frac{1}{4(x^{\pm})^2}$, where $x^{\pm}$ are
Kruskal coordinates, since this corresponds to vanishing
$t_{\pm}(\sigma^{\pm})$ in
null Minkowskian coordinates $\sigma^{\pm}$. In our case we should choose
$t_v(v)=t_u(u)=0$ and, according to (\ref{tr1}) and (\ref{tr2}), we
have\footnote{This type of behaviour for the functions $t_{\pm}$ was pointed out in
\cite{cfn}.}
\be
\label{boco}
t_+(x^+) = \frac{a^2 \Theta(x^+-x^+_0)}{(a^2-(x^+-x_0^+)^2)^2} \, .
\ee
Moreover, the coordinate $x^-$ always coincides with the null coordinate $u$ as
a consequence of the matching condition for the metric and dilaton with the
gauge fixing conditions (\ref{gauge+}) and (\ref{gauge-}).
\be
t_-(x^-) = 0 \, .
\ee
With the above choice, the constraints (\ref{const}) for the evaporating
solution ($x^+>x^+_0$) become
\bea
\partial^3_+ a_+ &=& -\frac{N\hbar}{12\pi} \frac{a^2}{(a^2-(x^+-x_0^+)^2)^2}
\, , \\
3\pmen a_- + 4x^- \partial^2_- a_- + (x^-)^2 \partial^3_- a_- &=&
\phantom{-} 0 \, ,
\eea
equations which can easily integrated to get the following solutions
\bea
a_+ &=& C(x^+ + \Delta^+) + \zeta_+ (x^+)^2 - \frac{N\hbar}{\pi} P(x^+) \, , \\
a_- &=& C(\frac{-2}{\l^2x^-} + \Delta^-) + \zeta_- (x^-)^{-2} \, ,
\eea
where $C$, $\zeta_{\pm}$ and $\Delta^{\pm}$ are integration constants and the
function $P(x^+)$ is
\be
\label{p}
P(x^+) = \frac{x^+-x_0^+}{48} - \frac{a^2 - (x^+-x_0^+)^2}{48a}
{\mathrm arctanh} \frac{x^+-x_0^+}{a} \, .
\ee
It is interesting to point out that the above function vanishes at $x^+_0$ and
the quantum solution is the classical one (\ref{aclass}) plus the correction
introduced through $P(x^+)$. This is so for $C=-\frac{1}{2}x_0^+\Delta m$,
$\zeta_{\pm}=0$ and $\Delta^{\pm}$ given by (\ref{d+}) and (\ref{d-}). The
dilaton function for $x^+>x_0^+$ is then
\be
\label{evaporating}
\tilde{\f} = lq^3 \frac{1-\frac{(x^+-x_0^+)(x^--x_0^+)}{a^2}}{x^--x^+} +
\frac{N\hbar}{2\pi} \frac{(x^--x^+) P^{\prime}(x^+) + 2P(x^+)}{x^--x^+} \, ,
\ee
which matches along $x^+=x_0^+$ with the extremal one, which continues being
solution at the semiclassical level
\be
\tilde{\f} = \frac{lq^3}{x^--x^+} \, .
\ee

A remarkable property of these solutions is that the dynamical
evolution can be followed analytically. As before, the curve $\tilde{\f}=0$
represents the location of the `extremal radius'
\be
\label{phizero}
x^- = \frac{lq^3 + \frac{lq^3x_0^+}{a^2} (x^+-x_0^+) -
\frac{N\hbar}{2\pi} x^+ P^{\prime} + \frac{N\hbar}{\pi} P}{\frac{lq^3}{a^2}
(x^+-x_0^+) - \frac{N\hbar}{2\pi} P^{\prime}} \, .
\ee
And for the apparent horizon $\partial_+\tilde{\f}=0$ in the spacetime of the
evaporating black hole, we get the following equation
\be
\label{apph}
lq^3 \left( 1 - \frac{(x^--x_0^+)^2}{a^2} \right) + \frac{N\hbar}{2\pi}
\left( \frac{1}{2} (x^--x^+)^2 P^{\prime \prime} + (x^--x^+) P^{\prime} +
P \right) = 0 \, .
\ee
The main property of both above curves is that, unlike the classical case and
unexpectedly, they intersect before reaching the AdS boundary at a finite
advanced time $x^+=x_{{\mathrm int}}^+$ given implicitly by the below relation
obtained by substituting (\ref{phizero}) in (\ref{apph})
\be
{\mathrm arctanh} \frac{x_{{\mathrm int}}^+-x_0^+}{a} = \frac{48lq^3}{Na} -
\frac{x_{{\mathrm int}}^+-x_0^+}{\sqrt{a^2-(x_{{\mathrm int}}^+-x_0^+)^2}}
\, .
\ee

\begin{figure}[H]
\centerline{\psfig{figure=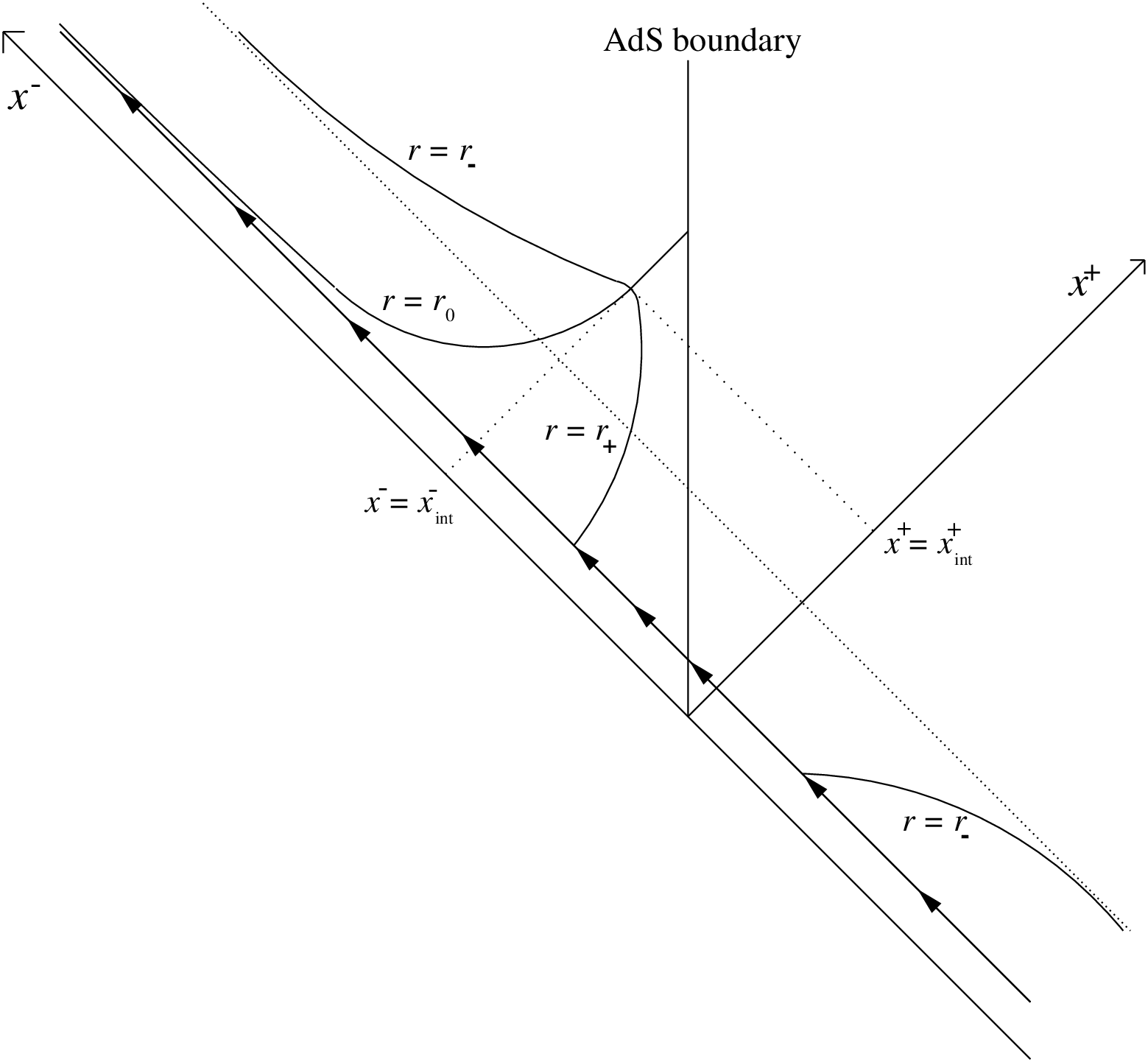,width=4.in,angle=-90}}
\end{figure}
\begin{center}
\makebox[12cm]{\parbox{12cm}
{\small Fig.3. Kruskal diagram of semiclassical evolution of
RN black hole. The outer apparent horizon shrinks until meets both inner
horizon and extremal radius curve at a finite advanced time. This is the
end-point of evaporation and a static solution can be matched so that the
extremal radius curve becomes the horizon of extremal black hole remnant.}}
\end{center}
\v1cm

A graphic description of this process is represented in Fig.3.
At this point the extremal radius curve $\tilde{\f}=0$ is null and both outer
and inner horizons meet. This means that we have arrived at the end-point of
the evaporation and, since we are dealing with analytic expressions, one can
check explicitly that the evaporating solution (\ref{evaporating}) matches
smoothly along $x^+=x_{{\mathrm int}}^+$ with a static solution for
$x^+>x_{{\mathrm int}}^+$
\be
\label{remnant}
\tilde{\f} = -k \frac{(x^+-x_{{\mathrm int}}^-)(x^--x_{{\mathrm int}}^-)}
{x^--x^+} \, ,
\ee
where
\be
k = \frac{\frac{lq^3}{a^2} (x_{{\mathrm int}}^+-x_0^+) - \frac{N\hbar}{2\pi} 
P^{\prime}(x_{{\mathrm int}}^+)}{x_{{\mathrm int}}^+ - x_{{\mathrm int}}^-}
\, ,
\ee
which turns out to be just the extremal black hole. A conformal coordinate
transformation brings the metric and $\tilde{\f}$ into the form (\ref{v-}) of
the extremal solution
\bea
\label{v3}
v &=& \frac{1}{\l^2k(x^+-x_{{\mathrm int}}^-)} \, , \\
\label{u3}
u &=& \frac{1}{\l^2k(x^--x_{{\mathrm int}}^-)} \, .
\eea

We must note that, in contrast with the studies of
dilatonic black holes \cite{rst1,rst2,bpp}, the matching of the evaporating
solution here is along the line $x^+=x_{{\mathrm int}}^+$ and not
$x^-=x_{{\mathrm int}}^-$. There is a physical reason for this, the line
$x^+=x_{{\mathrm int}}^+$ in our effective theory represents, as we have
stressed before, points very near to the apparent horizon of the 
evaporating RN black hole. Moreover, the matching is smooth and $T^f_{++}$
vanishes at $x_{{\mathrm int}}^+$ in contrast with the dilatonic black holes,
for which there is an emanating thunderpop of negative energy
\cite{rst1,rst2}.\\

Concerning the mass curve for the evaporating solution (\ref{evaporating}) in
$x_0^+ < x^+ < x_{{\mathrm int}}^+$, it corresponds to a chiral energy
distribution which accounts for the neutral mass lost by the near-extremal
black hole during its evaporation. Taking into account (\ref{localmass}) we
get
\bea
\label{lmass}
\tilde{m}(x^+) &=& \Delta m - \frac{N\hbar}{12\pi a} {\mathrm arctanh}
\frac{x^+-x_0^+}{a} - \nonumber \\
&& \frac{N^2\hbar^2}{1152\pi^2 lq^3} \left[ \frac{(x^+-x_0^+)^2}
{a^2-(x^+-x_0^+)^2} - {\mathrm arctanh}^2 \frac{x^+-x_0^+}{a} \right] \, ,
\eea
that just vanishes at the end-point $x^+=x_{{\mathrm int}}^+$ as it is showed
in Fig.4.\\

To end this section we would like to comment on the boundary condition
(\ref{boco}). It was derived through the Schwarzian derivative of the classical
relations (\ref{tr1})), (\ref{tr2}). This should be considered as a first
approximation since the evaporating solution modifies the classical relation
(\ref{tr2}). Plugging (\ref{lmass}) and (\ref{evaporating}) into (\ref{rn2h})
we get the following relation to leading order in $\hbar$
\be
v = x_0^+ + a \> {\mathrm arctanh} \frac{x^+-x_0^+}{a} -
\frac{a^2N\hbar}{96lq^3} \left( \frac{a^2}{a^2-(x^+-x_0^+)^2}-
{\mathrm arctanh}^2 \frac{x^+-x_0^+}{a} \right) \, .
\ee
As a consequence of this, the function $t_+(x^+)$ get modified
\be
t_{x^+} = \frac{a^2}{(a^2-(x^+-x_0^+)^2)^2} - \frac{a^4N\hbar}{24lq^3}
\frac{x^+-x_0^+}{(a^2-(x^+-x_0^+)^2)^3} \, .
\ee
We should remark that this quantum correction for the function $t_+(x^+)$ does
not happen in the CGHS model, due to the fact that the temperature is
independent of the mass. In our case the back reaction is more involved and
produces this type of effect. Solving the equations (\ref{qeq1}-\ref{qeq4})
again in terms of the quantum corrected function $t_+(x^+)$ ($t_-(x^-)$ remains
zero) and repeating the process described in this section we arrive to a new
dynamical function $\tilde{\f}$ and evaporating mass $\tilde{m}(x^+)$
describing the evaporation. But what one surprisingly finds is an unphysical
evolution with periods of antievaporation. This suggests that only an exact
treatment of the boundary functions $t_+(x^+)$ can produce a correct result.
This will be considered in the next section.

\v1cm
\begin{figure}[H]
\centerline{\epsfig{figure=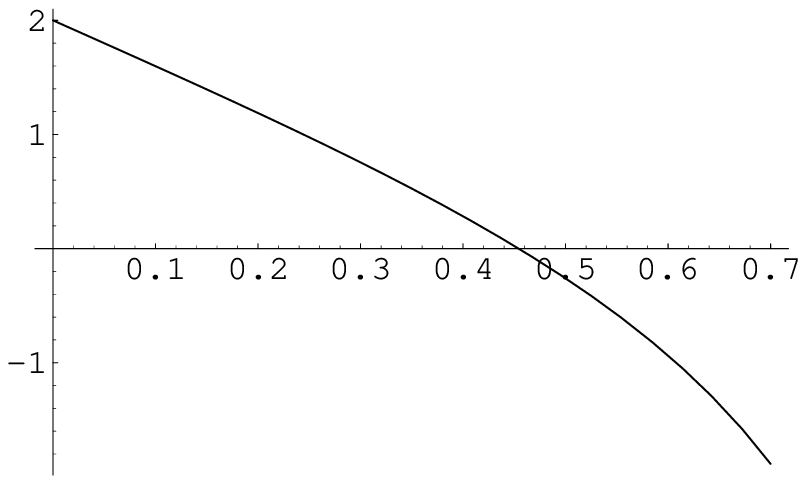,width=3in,angle=0}}
\end{figure}
\begin{center}
\makebox[12cm]{\parbox{12cm}
{\small Fig.4. Dynamical evolution of the mass deviation from extremality of
the evaporating near-extremal black hole. We set $x_0^+=0$, $\Delta m=2$,
$l=q=1$ and $N=48$ in (\ref{lmass}). It vanishes at $x_{{\mathrm int}}^+$
indicating the end-point of the evaporation. Even though the evaporation could
seem to continue for negative values of the mass, the static extremal solution
can be matched at this point.}}
\end{center}
\v1cm

\setcounter{equation}{0}
\section{Back reaction for dynamical $t_{\pm}$ functions}

We shall now modify the scheme used in the previous section to work out the
gravitational back reaction. The crucial point is to consider the functions
$t_{\pm}(x^{\pm})$, appearing in the equations (\ref{qeq4}), as dynamical
objects to be determined at the end together with the solutions for the dilaton
and the metric. We should do it requiring the physical consistence of the
procedure. To start with, it is clear that $t_-(x^-)$ does not get any quantum
correction since the coordinate $x^-$ remains equal to the $u$ coordinate of
the extremal solution. Therefore we have
\be
\label{t6}
t_-(x^-) = 0 \, .
\ee
The advantage of (\ref{t6}) is that, irrespective to the form of $t_+(x^+)$,
the general solutions for the metric (even in the presence of arbitrary
incoming classical matter $T^f_{++}$) is given by \cite{cnn}
\be
\label{rnevaporating}
ds^2 = -\left( \frac{2\tilde{x}^2}{l^2q^3}-l\tilde{m}(v) \right) dv^2 +
2d\tilde{x} dv \, ,
\ee
where $\tilde{m}(v)=m-m_0$ is the deviation of evaporating mass from
extremality. This metric can be brought into the gauge-fixed form
(\ref{jt1}) by means of the transformation
\bea
v &=& v(x^+) \, , \\
\tilde{x} &=& l \tilde{\f} (x^{\pm}) \, ,
\eea
where
\be
\frac{dv}{dx^+} = \frac{-lq^3}{\pmen \tilde{\f} (x^--x^+)^2} \, .
\ee
This requires that $\pmen \tilde{\f} (x^--x^+)^2$ is a chiral function of
$x^+$ and this follows from the constrain equation
\be
-2\pmen^2 \tilde{\f} + 4 \pmen \r \pmen \tilde{\f} = 0 \, .
\ee
Since in the advanced time coordinate $v$ the function $t_v(v)$ vanishes, in
the coordinate $x^+$ it is
\be
\label{bc1}
t_+(x^+) = \frac{1}{2} \{ v, x^+ \} \, .
\ee
Now we find convenient to introduce the function $F(x^+)$ such that
\be
\label{vchange}
\frac{dv}{dx^+} = \frac{lq^3}{F(x^+)} \, .
\ee
In gauge (\ref{gauge+}), (\ref{gauge-}), the equations
(\ref{qeq1}-\ref{qeq4}) can be integrated leading to 
\be
\label{evphi}
\tilde{\phi} = \frac{F(x^+)}{x^--x^+} + \frac{1}{2}  F'(x^+) \, ,
\ee
where the function $F(x^+)$ satisfies the following differential equation
\be
\label{diffequation}
F'''= \frac{N\hbar}{24\pi} \left( -\frac{F''}{F} +
\frac{1}{2} \left( \frac{F'}{F} \right)^2 \right) \, .
\ee
Physical considerations concerning
the matching along $x^+=x^+_0$ between the extremal and the evaporating
solutions provide the boundary conditions for the above differential equation.
Namely, from the continuity of the function $\tilde{\f}$ (\ref{evphi})
along $x^+_0$ we get
\be
F(x_0^+) = lq^3 \, , \qquad F^{\prime}(x_0^+) = 0 \, ,
\ee
whereas, from the discontinuity of the shock-wave stress tensor $T^f_{++}=
\Delta m \delta (x^+-x^+_0)$, we obtain
\be
F^{\prime \prime} (x_0^+) = -\Delta m \, .
\ee
It is not difficult to relate the deviation of the evaporating mass
$\tilde{m}(x^+)$ with the function $F(x^+)$
\be
\label{exactmass}
\tilde{m}(x^+) = \frac{24\pi }{lq^3 N\hbar} F^2  F''' \, .
\ee
Moreover the boundary function $t_+(x^+)$ (\ref{bc1}) can be written as
\be
\label{tp}
t_+(x^+) = \frac{lq^3\tilde{m}(x^+)}{2F^2} \, ,
\ee
which makes clear the dynamical content of the function $t_+(x^+)$. To compute
the evaporating mass $\tilde{m}(x^+)$ we need to solve equation
(\ref{diffequation}) but, fortunately, in terms of the advanced time coordinate
the problem is simpler. Derivating (\ref{exactmass}) with respect to $x^+$,
one can readily arrive to
\be
\label{massx}
\pmas \tilde{m}(x^+)=-\frac{N\hbar}{24\pi lq^3}\frac{\tilde{m}(x^+)}{F} \, ,
\ee
which can be readily integrated in coordinate $v$
\be
\label{massv}
\tilde{m}(v)=\Delta m e^{-\frac{N\hbar}{24\pi lq^3}(v-v_0)} \, ,
\ee
leading to an infinite evaporation time.

\v1cm
\begin{figure}[H]
\centerline{\psfig{figure=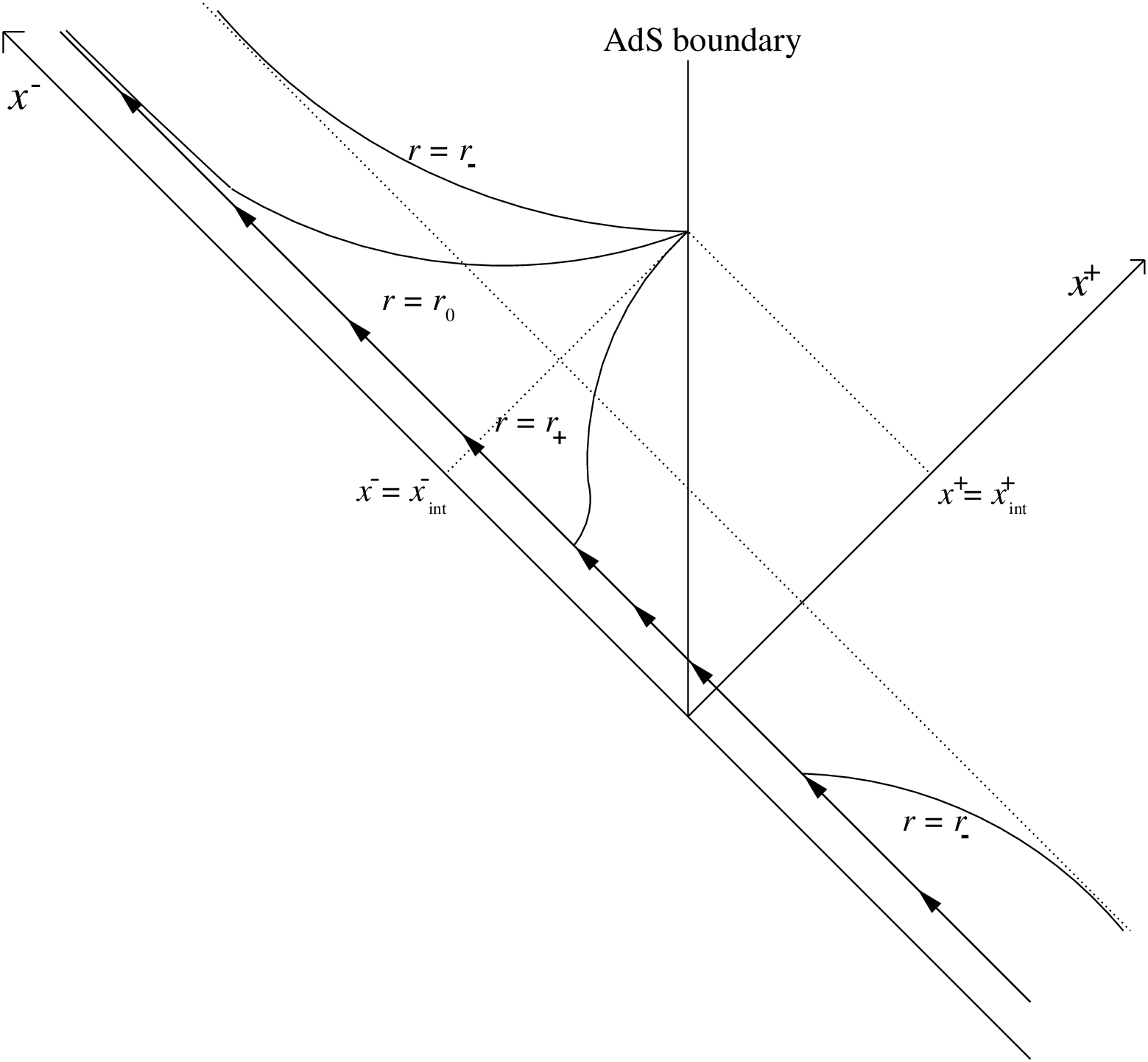,width=4.in,angle=-90}}
\end{figure}
\begin{center}
\makebox[12cm]{\parbox{12cm}
{\small Fig.5. Kruskal diagram of semiclassical evolution of
RN black hole. The outer apparent horizon shrinks until it meets both inner
horizon and extremal radius curve at the AdS boundary.}}
\end{center}
\v1cm

Now let us go back to the $x^{\pm}$
frame. As seen in last section, the physical information about the dynamical
evolution of the near-extremal black hole is encoded in the function
(\ref{evphi}). The evolution of the `extremal radius' is represented by the
curve $\tilde{\f}=0$
\be
\label{aor}
x^-=x^+ - \frac{2F}{F'} \, ,
\ee
and the curve $\partial_+ \tilde{\f} =0$ accounts for the inner and outer
apparent horizons
\be
\label{por}
x^-=x^+ -\frac{F'}{F''} \pm\frac{\sqrt{F'^2 -2F F''}}{F''} \, .
\ee 
These three curves meet only when 
\be
\label{ep}
F'(x^+_{int})^2 - 2F(x^+_{int}) F''(x^+_{int}) = 2lq^3m(x^+_{int}) = 0 \, ,
\ee
and it takes place at a finite value $x^+_{int}$ at the end of the
evaporation. Since $v \rightarrow +\infty$ as $x^+ \rightarrow x^+_{int}$,
taking into account (\ref{vchange}) we also get that $F(x^+_{int})=0$. And from
this feature it follows immediately, see (\ref{aor}) and (\ref{por}), that 
$x^+_{int}=x^-_{int}$ and then the intersection point belongs to the AdS
boundary so that it gives an infinite amount of proper time in accordance with
(\ref{massv}). We can also show that
$F'''(x^+_{int})=F'(x^+_{int})=0$ and since for all of these three curves
(\ref{aor}), (\ref{por}), we have
\be
\frac{dx^-}{dx^+} \stackrel{x^+ \rightarrow x^+_{int}}{\longrightarrow}
0 \, ,
\ee
one can conclude that the three curves meet at the end-point becoming a null
line. The complete physical process is represented in Fig.5.\\

To finish this analysis we consider the numerical solution to the differential
equation (\ref{diffequation}). From the numerical plot of the function
$F(x^+)$ and its derivatives (see Fig.6a), apart from the properties found
before ($F(x^+_{int})=F'(x^+_{int})=F'''(x^+_{int})=0$), it also follows that
that $F''(x^+_{int})$ is nonzero while further derivatives vanish. Thus
locally close to the intersection point $F(x^+)$ behaves as a parabola with 
exponentially suppressed corrections. The numerical plot of $\tilde{\f} =0$
and $\partial_+ \tilde{\f} =0$ coincides exactly with that of Fig.5. The
saddle point in  the outer apparent horizon curve $r_+$ signals the transition
from the strong to the weak back-reaction regimes as discussed in \cite{lo}.
At the end-point the curves $\tilde{\f}=0$ and $\partial_+\tilde{\f}=0$ are
null and the
dilaton function is well represented asymptotically by the extremal solution
\be 
\tilde{\f} = \frac{F^{\prime \prime}(x^+_{int})}{2}
\frac{(x^+-x_{{\mathrm int}}^+)(x^--x_{{\mathrm int}}^+)}{x^--x^+} +
{\cal{O}}(e) \, ,
\ee
where ${\cal{O}}(e)$ are exponentially small high order terms. So the
evaporating black hole approaches asymptotically to the extremal configuration
without actually coming back to the extremal state in a finite continuous
process. This result appears to be well motivated from a thermodynamical point
of view, in particular from Stefan's law  $\frac{dm}{dt}\sim -4\pi r_+^2 T_H^4$
which predicts that a near-extremal evaporating black hole does never come back
to the extremal state \cite{ddr}. This feature is closely related to Nernst's
version of
the third law of thermodynamics which states that the temperature of a system
cannot be reduced to zero in a finite number of operations. Israel \cite{isr}
showed that the same conclusions must be true in the case of the RN black hole
provided that the stress energy tensor of the infalling matter satisfies the
weak energy condition (WEC). It is well known that in the Hawking process the
WEC is violated close to the horizon; nevertheless our exact results do not
violate the third law. We think that this is an encouraging sign towards
generalising its validity to more general (quantum) frameworks.

\v1cm
\begin{figure}[H]
\centerline{\epsfig{figure=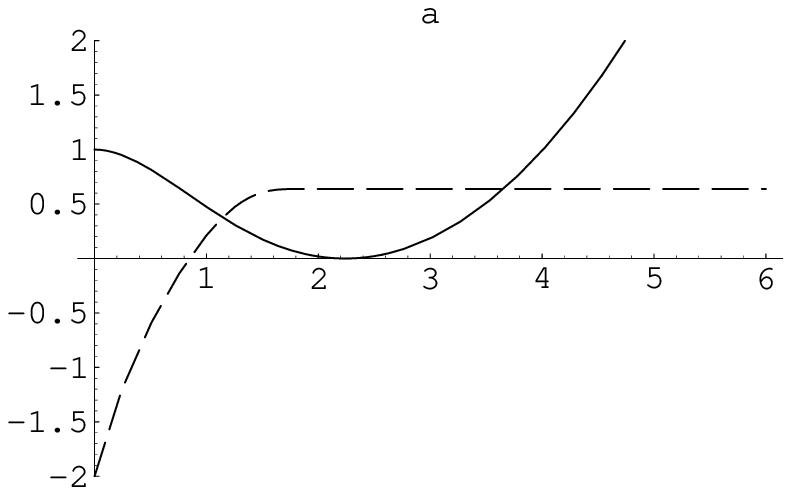,width=2.5in,angle=0}
\epsfig{figure=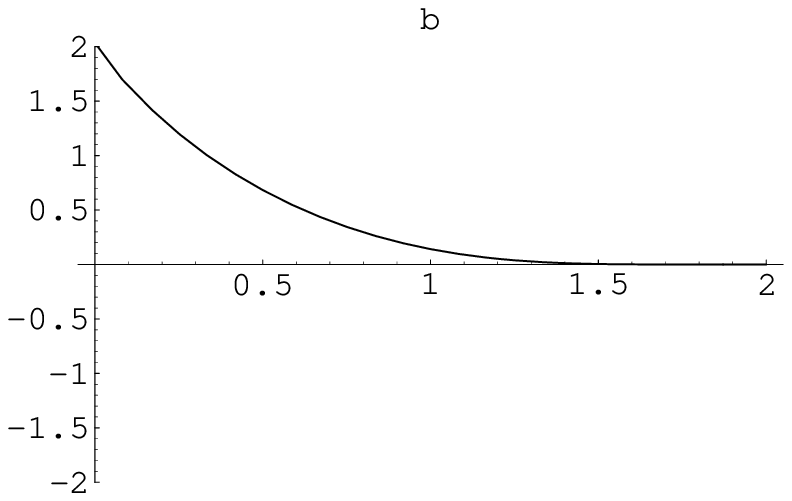,width=2.5in,angle=0}}
\end{figure}
\begin{center}
\makebox[12cm]{\parbox{12cm}
{\small Fig.6. Numerical plots for some solutions to the differential equation
(\ref{diffequation}), $x_0^+=0$, $\Delta m=2$, $l=q=1$ and $N=48$. (a) Plot of
the function $F(x^+)$ (continuous line) and its second derivative (dashed
line). The zero of $F$ corresponds to the value $x^+_{int}$ while the second
derivative is non-zero at this point. (b) Plot of the dynamical evolution of
the mass deviation. It approaches smoothly to zero as it reaches $x^+_{int}$.}}
\end{center}
\v1cm

\setcounter{equation}{0}
\section{Conclusions and final comments}

In this paper we have studied an effective model which describes the physics
of near-extremal RN black holes in a region very close to the horizon. We have 
focused our analysis on the evaporation process produced when a low-energy
shock wave excites the extremal black hole and the non-extremal configuration
decays via Hawking emission. We have shown that a self-consistent treatment
of the gravitational back reaction requires to consider the non-local
contributions $t_+(x^+)$ of the effective action as dynamical variables. So
doing this we find a physical picture of the evaporation process compatible
with the third law of thermodynamics. An infinite amount of time is necessary
for the black hole to go back to extremality and this `happens' in the $AdS_2$
boundary of the near horizon geometry. We have to remark that we have obtained
a rather accurate description near the horizon. For the in-falling observer
the function $t_+(x^+)$ (\ref{tp}) is proportional to the flux of negative
energy across the horizon and, therefore, responsible of the black hole
radiation. However our scheme do not describe directly the effects of the back
reaction on the radiation measured by an asymptotic observer (not very close to
the horizon). This makes our results compatible with the principle of
complementarity \cite{th,stu,svv}.\\

Finally we would like to comment on the implications of considering a
continuous distribution of incoming matter $T^f_{vv}$. The equation
(\ref{massx}) in terms of the $v$ coordinate is then
\be
\label{masseq}
\partial_v \tilde{m}(v) = -\frac{N\hbar}{24\pi lq^3} \tilde{m}(v) +
T^f_{vv} \, .
\ee
We observe that the evaporating mass $\tilde{m}(v)$, and hence the function
$t_+(x^+)$, contains detailed information of the classical matter. In other
words, the information of the stress tensor $T^f_{vv}$ is also codified in the
quantum incoming flux\footnote{This is not possible for the CGHS theory since
the temperature is independent of the mass.}
\be
-\frac{N\hbar}{12\pi} t_+(x^+) \, .
\ee
Note that this is true because all the higher-order quantum corrections to
$t_+(x^+)$ have been taken into account, otherwise we could not get
(\ref{masseq}). In the approximation of section 5 the information is lost. The
functions $t_+(x^+)$, at leading order, does not see the details of $T^f_{vv}$.
Moreover the full solution, in contrast with the models \cite{rst1,rst2,bpp},
seems
to depend on all higher-order momenta of the classical stress tensor. All this
means that the information might not be lost. However to get definite
conclusions we should be able to describe the outgoing radiation at infinity
and this is out of the reach of the near-horizon scheme of this paper.
Nevertheless energy conservation suggests that an analogous mechanism to that
storing the information in $t_+(x^+)$ across the horizon should radiate the
information out to infinity.

\section*{Acknowledgements}

This research has been partially supported by the CICYT and DGICYT, Spain.
D. J. Navarro acknowledges the Ministerio de Educaci\'on y Cultura for a FPI
fellowship. A.F. thanks R. Balbinot for useful discussions. D.J.N. and J.N-S.
also wish to thank J. Cruz and P. Navarro for comments.



\end{document}